\documentclass[preprint, 12pt]{elsarticle}
\usepackage{graphicx} 
\usepackage{svg}
\usepackage{amsmath}
\usepackage{url}
\usepackage{subfigure}

\usepackage{amssymb}
\DeclareMathAlphabet{\mathbbold}{U}{bbold}{m}{n}

\begin{document}
\begin{frontmatter}

\title{The Marked Edge Walk: A Novel MCMC Algorithm for Sampling of Graph Partitions}

\author[1]{Atticus McWhorter\corref{cor1}}
\affiliation[1]{
  organization={Department of Mathematics, Dartmouth College},
  addressline={1 College Street},
  city={Hanover},
  postcode={03755},
  state={NH},
  country={USA}
}
\ead{atticus.w.mcwhorter.gr@dartmouth.edu}

\author[2]{Daryl DeFord}
\ead{ddeford@vassar.edu} 
\cortext[cor1]{Corresponding author}

\affiliation[2]{
  organization={Department of Mathematics and Statistics, Vassar College},
  addressline={124 Raymond Avenue},
  city={Poughkeepsie},
  postcode={12604},
  state={NY},
  country={USA}
}

\begin{abstract} 
    Novel Markov Chain Monte Carlo (MCMC) methods have enabled the generation of large ensembles of redistricting plans modeled  as a graph partitioning problem. However, existing algorithms such as Reversible Recombination (RevReCom) and Metropolized Forest Recombination (MFR) have strong preferences for distributions related to the spanning tree measure. In this paper we introduce the Marked Edge Walk (MEW), a novel Markov chain proposal for sampling from the space of graph partitions. The walk operates on the space of spanning trees with marked edges, allowing for calculable transition probabilities for use in the Metropolis-Hastings algorithm. Empirical results on real-world dual graphs show convergence under a broad class of target distributions less constrained by spanning tree counts, including policy-based distributions, such as competitiveness on New Hampshire that are independent of spanning trees, and compactness and partisan symmetry distributions on New Hampshire and Texas that, while related to spanning trees, can now be properly targeted with a smaller degree of spanning tree bias, which represents an advancement in flexible ensemble generation.
\end{abstract}


\end{frontmatter}

\section{Introduction}
Recent advances in computational capabilities and data availability have greatly increased legislators' abilities to optimize political redistricting plans. In these algorithms, US states are partitioned into geographic units which are then encoded as planar graphs, transforming the problem to one about studying connected graph partitions \citep{deford2019recombination}. Quantitative analysis of redistricting plans in this setting has often been carried out by generating large `ensembles' of comparator plans using random walks to explore the space of balanced graph partitions \citep{aftermath}. In these efforts to develop a baseline of `typical' plans, researchers have largely turned to Markov Chain Monte Carlo (MCMC) \citep{deford2019redistricting, colorado, caldera2020mathematics, deford_compete, chikina_assessing_2017, herschlag_quantifying_2018, Fifield_A_2018} or other Sequential Monte Carlo (SMC) methods \citep{mccartan2023sequential, osullivan2026generalizedsequentialmontecarlo} to generate large collections of plans based on parameters drawn from legislative guidance or historical plans.

\medskip

Before discussing specific algorithms, it is worth motivating why uniform sampling from the space of redistricting plans is not our goal. As discussed in \citet{najt2020empirical}, an unconstrained uniform distribution is not a useful baseline for policy analysis because most plans in the space are extremely non-compact, and the uniform distribution assigns equal weight to highly irregular plans. One approach to counteract this might be to restrict the state space with binary constraints on relevant criteria, for example by excluding all partitions that are significantly imbalanced or non-compact. This runs the risk of disconnecting the state space and also does not preferentially sample from plans that score better on relevant criteria, which is a desirable feature in applied work in this area. This type of approach has been used effectively when combined with some of the other methods described below \citep{herschlag_quantifying_2018}.

\medskip

Determining the distributions over graph partitions that can be sampled from efficiently and effectively is an active area of research. There are generic hardness results for uniform sampling of connected, balanced partitions on planar graphs \citep{najt2019complexity}, and a developing literature on better bounds for uniform sampling grid-like graphs \citep{frieze2022subexponentialmixingpartitionchains}, subexponential algorithms \citep{cai2024uniformlyrandomsolutionalgorithmic}, and the corresponding enumeration problem \citep{donnay2024asymptoticsredistrictingntimesn}. Beyond the uniform distribution, there has also been a significant amount of recent theoretical work on sampling from the spanning tree distribution which is described in more detail below \citep{cannon2022spanning, cannon2024samplingbalancedforestsgrids, cannon2025samplingtreeweightedpartitionssampling, charikar2023complexitysamplingredistrictingplans}. 
Given that the uniform distribution is often an undesirable target and that the governing legislation is not described in terms of the combinatorial spanning tree measurements that many of the current methods are most effective at sampling from, this motivates our interest in developing algorithms that can target a broader class of distributions. 


\medskip

A number of different algorithms are available to researchers who hope to generate an ensemble of redistricting plans, including several with public open-source implementations. As a developing field however, there is still room for improvement in methodology. A natural approach is to generalize proposals from statistical physics models, making small updates to the partition labels at each step \citep{chikina_assessing_2017, Chikina_SPP, Fifield_A_2018}.
The Recombination (ReCom) algorithm \citep{DeFord2021Recombination}, which iteratively merges pairs of adjacent districts and then bipartitions the merged pair by sampling a spanning tree and cutting an edge, has gained significant popularity both in academic work \citep{caldera2020mathematics, deford_compete, deford2019redistricting, colorado, NHGerry} and in litigation, due to its favorable mixing properties, and studies using ReCom have been used to argue the (un)constitutionality of districting plans in courts as high as the U.S. Supreme Court \citep{mathematiciansbrief}. However, the invariant distribution of ReCom is unknown \citep{cannon2022spanning}.

\medskip

The Reversible Recombination (RevReCom) algorithm \citep{cannon2022spanning} tailors ReCom to satisfy the detailed balance condition and target the spanning tree distribution, a distribution which weights each graph partition according to the product of the number of spanning trees of each part. Great attention has been devoted to the spanning tree distribution \citep{cannon2022spanning, autry2021metropolized, autry2020multiscalemergesplitmarkovchain, charikar2023complexitysamplingredistrictingplans, clelland2021compactnessstatisticsspanningtree}. \citet{Procaccia2021Compact} showed that the spanning tree distribution is inversely exponentially related to the total boundary lengths of the districts, and \citet{cannon2024samplingbalancedforestsgrids} developed an algorithm that can sample from the distribution in polynomial time. 

\medskip

Despite great attention to the spanning tree distribution, there is still interest in sampling from more general distributions, particularly those that incorporate policy considerations from a given state. Metropolized Forest Recombination (MFR) \citep{autry2023metropolized} is designed to do exactly that, utilizing the powerful tools of the Metropolis-Hastings algorithm to target any desired invariant distribution. To achieve this, the authors `lift' to an expanded state space of spanning forests that allows for easier calculation of the forward and reverse transition probabilities. However, it is still an open question what distributions can effectively be sampled using this method. As with ReCom, this approach has been successfully applied in both academic contexts \citep{zhao2022mathematically} as well as litigation, and it has also been extended to the multiscale setting \citep{autry2020multiscale, chuang2024multiscaleparalleltemperingfast}. 

\medskip

One advantage of the MFR approach presented by \citet{autry2023metropolized} is that it makes explicit the contribution of the spanning tree density to the target distribution with a parameter that interpolates the relative impact of this term. We discuss this in more detail in Section \ref{sec:target} below, but a key motivation for our work is targeting distributions that are not dominated by this term. This motivation is common to other recently proposed sampling methods that take a similar approach to the one we develop here, in walking on the lifted space of trees in an attempt to reduce the reliance on the spanning tree distribution. In particular, the Cycle Walk \citep{deford2025cyclewalksamplingmeasures} and Balanced Up Down (BUD) Walk \citep{akitaya2026balancedupdownwalk} fall into this category. Section 3.1 of  \citep{akitaya2026balancedupdownwalk} describes the relationships and motivations between these proposals and the one we present here. 

\medskip

In this paper, we present the marked edge walk (MEW), an algorithm for sampling from the space of balanced graph partitions under a tuneable distribution. We demonstrate convergence on the uniform distribution on Cheshire County, competitiveness, compactness, and multivariate distributions on New Hampshire, and a multivariate distribution on Texas. We note that when targeting a Gaussian distribution on cut edges, MEW implicitly targets a lognormal distribution on spanning trees. While this remains dependent on spanning trees, it still represents an advancement over prior methods, which only partially account for the spanning tree degeneracy factor.

\section{Procedure}
\label{sec:Procedure}

The space of interest of MEW is the space of balanced graph partitions of a fixed graph $G = (V, E)$. In our application, we choose $G$ to have a finite set of vertices $V$ and undirected edges $E$. A graph partition $\xi : V \rightarrow \{1, 2, ..., d\}$ sorts each vertex into a part. However, instead of representing each state as a graph partition, as  \citet{DeFord2021Recombination}, we instead lift to the space of spanning trees with marked edges. More formally, the state space of MEW is \begin{equation}
    \mathcal{X} = \{(T, M): T  \text{ is a spanning tree of } G, M \subseteq T\}.
\end{equation} 
\medskip

We will enforce that the forest $F = T \setminus M$ is balanced, where a forest $F$ is balanced if each connected component $D_i$ satisfies $|D_i| = n/d \pm \epsilon$ (node balance) or $\sum_{v \in D_i} w(v) = W/d \pm \epsilon$ (population balance) for vertex weights $w: V \rightarrow \mathbb{R}_{\geq 0}$ and total weight $W = \sum_{v \in V} w(v)$. In this work, we enforce population balance by rejecting proposals that yield unbalanced forests, similar to the approach in \citet{DeFord2021Recombination}. Another approach can be found in \citet{autry2023metropolized, Autry_Multiscale}, where the authors incorporate balance conditions in the target distribution, allowing for both hard and soft balance constraints.

\medskip 

From a state $x \in \mathcal{X}$, the MEW transition consists of two parts: 

\begin{enumerate}
    \item \textbf{Cycle basis step:} Choose an edge $e_+ \in E \setminus T$ according to some distribution $\nu_1$. Let $C$ be the unique cycle in $T \cup \{e_+\}$. Choose an edge $e_- \in C$ according to some distribution $\nu_2$ and subject to the constraint that $e_- \notin M$ (marked edges cannot be removed), and set $T' = (T \cup \{e_+\}) \setminus \{e_-\}$.
    \item \textbf{Marked edge step:} Choose a marked edge $m \in M$ according to some distribution $\mu_1$. Choose an endpoint $u \in m$ according to some distribution $\mu_2$. Choose a neighbor $v \in N_{T'} (u)$ according to some distribution $\mu_3$,  and let $m' = \{u, v\}$. Set $M' = (M \cup \{m'\}) \setminus \{m\}$.
\end{enumerate}

\medskip

\begin{figure}[htbp]
       \centering{
\renewcommand{\arraystretch}{0.01}
        \begin{tabular}{cc}
        \includegraphics[width=0.49\linewidth]{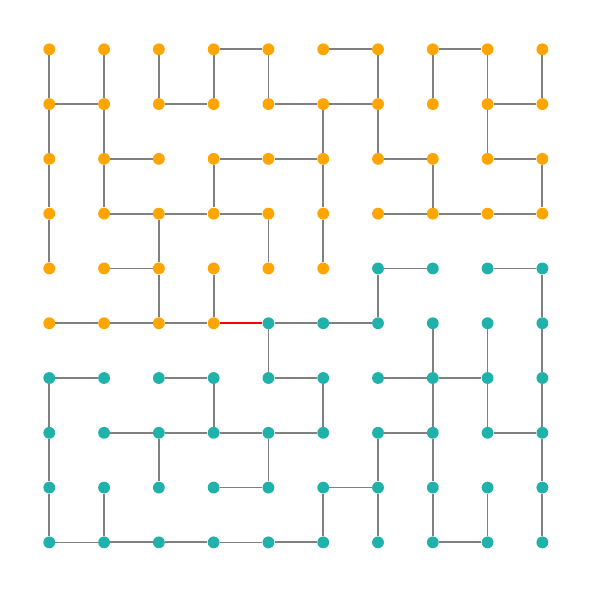}  &  \includegraphics[width=0.49\linewidth]{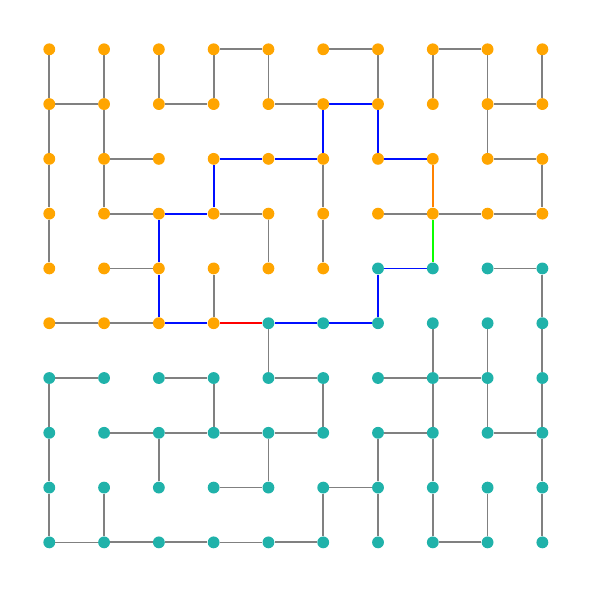} \bigskip \\
        \includegraphics[width=0.49\linewidth]{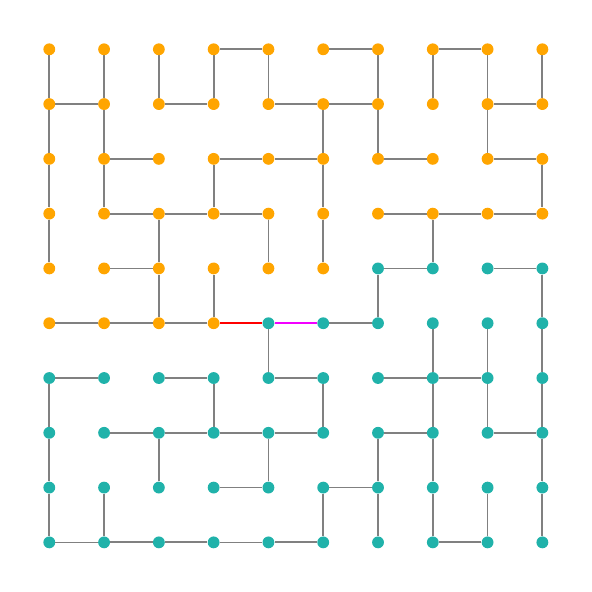}  &  \includegraphics[width=0.49\linewidth]{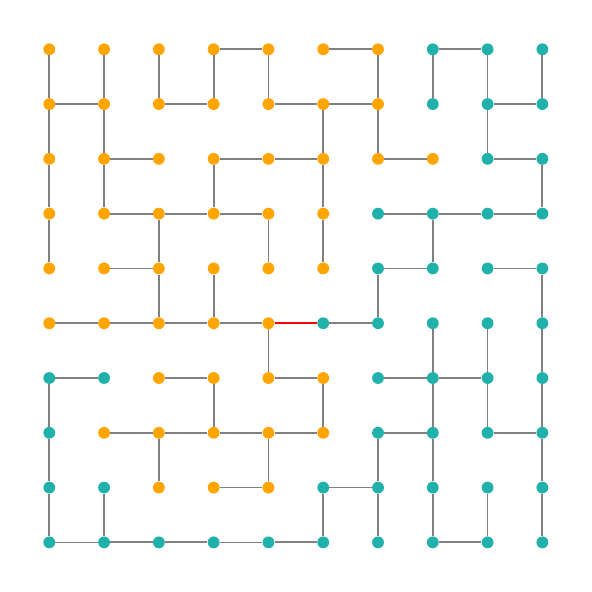}
        \end{tabular}
       }
       \caption{{\footnotesize  Illustration of MEW's two-step procedure. (Top-left) Initial state with spanning tree (black edges) and marked edges (red) that create a balanced partition when removed. (Top-right) Cycle basis step: adding edge $e_+$ (green) creates cycle $C$ (blue), then removing edge $e_-$ (orange) yields new tree $T'$. (Bottom-left) Marked edge step: the right vertex of the marked edge (red) is chosen, and its right neighbor is chosen, yielding a new marked edge (pink). (Bottom-right) These two steps create a new partition, completing one step of the MEW procedure. }
       \label{fig:scheme}}
\end{figure}

The choice to select a neighbor of an endpoint of the current marked edge, rather than an arbitrary edge from $T'$, is deliberate. By moving the marked edge locally in the tree, the resulting partition is more likely to remain balanced, reducing the rate of rejections due to population imbalance. For the current study and the following calculations, we choose $\nu_1, \nu_2, \mu_1, \mu_2,$ and $\mu_3$ to be uniform, but one could easily incorporate other distributions to, for example, avoid repeated selection of high-degree vertices, incentivize the selection of plans with low isoperimetric scores, or disincentivize the splitting of counties or towns as in  \citet{colorado, NHGerry}. A schematic of the procedure as chosen in our study is shown in Figure \ref{fig:scheme}. 

\medskip

An alternate formulation of MEW could take a probabilistic choice between the cycle basis step and the marked edge step. In this variant, with probability $p$, one would take a cycle basis step, and with probability $1 - p$, a marked edge step. This approach has the benefit of symmetric transitions, adding computational simplicity and yielding the theoretical result that, if the walk is irreducible, the unique stationary distribution is the uniform distribution on $\mathcal{X}$. The cycle basis step can be viewed as an up-down step as in the recently proposed BUD Walk \citep{akitaya2026balancedupdownwalk} or a 1-cycle step in the cycle walk \citep{deford2025cyclewalksamplingmeasures}.

\section{Transition Probabilities}

While the probabilistic variant targets the uniform distribution, for redistricting applications, we often want to target alternative distributions $p(x)$. To achieve this, we employ MEW (performing both a cycle basis step and a marked edge step sequentially) as a proposal in the Metropolis-Hastings algorithm. The algorithm generates samples from a target distribution $p(x)$ by rejection sampling according to an acceptance probability \begin{equation}
    a = \min \left\{1, \frac{p(x')}{p(x)}\frac{P(x|x')}{P(x'|x)}\right\},
\end{equation} where $P(x'|x)$ and $P(x|x')$ are the forward and backward transition probabilities of MEW. 

\medskip

Given two states $x = (T, M)$ and $x' = (T', M')$ (as defined in Section \ref{sec:Procedure}), the transition probability $P(x'|x)$ decomposes naturally according to the two components of the tuple $(T, M)$. In this section, we present the transition probabilities under uniform selection distributions $\nu_1, \nu_2, \mu_1, \mu_2,$ and $\mu_3$. The full transition probability for general selection distributions is tractable but lengthy and is therefore omitted here. 

\medskip

The tree component of the transition probability depends on the length of the cycle $C$ formed by $T \cup \{e_+\}$, and the constraint that marked edges cannot be removed. The forward transition probability $P(T'|T, M) = (|E \setminus T|\cdot|C \setminus M|)^{-1}$, that is, the probability of choosing the correct edges $e_+$ and $e_-$. Taking the ratio, \begin{equation}
    \frac{P(T|T', M')}{P(T'|T, M)} = \mathbbold{1}(m' \neq e_+) \cdot \frac{|C \setminus M|}{|C \setminus M'|},
\end{equation} where $\mathbbold{1}(X)$ is the indicator function that returns 1 if X is true and 0 otherwise. This term is needed because if the moved marked edge $m' = e_+$, then $P(T|T') = 0$ (since we cannot remove a marked edge). 

\medskip

The marked edge transition probability depends on the vertex degrees in the spanning tree and whether the marked edge configuration changes. Let $\deg_G(u)$ refer to the degree of vertex $u$ in the graph $G$. Then the forward transition probability is $P(M'|T', M) = 1/|M| \cdot 1/2 \cdot (1/\deg_{T'}(u) + \mathbbold{1}(m = m')/\deg_{T'}(v))$, that is, the probability of choosing the correct marked edge $m$, endpoint $u$, and neighbor $v$. The indicator term accounts for the possibility that when $m = m'$ (the marked edge doesn't move), either endpoint could be selected. Taking the ratio, \begin{equation}
    \frac{P(M|T, M')}{P(M'|T', M)} = \frac{\deg_{T'}(u)}{\deg_T(u)} \cdot\left[\frac{\deg_T(u) + \deg_T(v)}{\deg_{T'}(u) + \deg_{T'}(v)} \cdot\frac{\deg_{T'}(v)}{\deg_{T}(v)}\right]^{\mathbbold{1}(m = m')}.
\end{equation}

\medskip
Putting it all together, \begin{equation}
    \frac{P(x|x')}{P(x'|x)} = \mathbbold{1}(m' \neq e_+) \cdot \frac{|C \setminus M|}{|C \setminus M'|} \cdot \frac{\deg_{T'}(u)}{\deg_T(u)} \cdot \left[\frac{\deg_T(u) + \deg_T(v)}{\deg_{T'}(u) + \deg_{T'}(v)} \cdot\frac{\deg_{T'}(v)}{\deg_{T}(v)}\right]^{\mathbbold{1}(m = m')}.
\end{equation}
\medskip

In most cases, this expression will simplify to unity because most cycles do not intersect the changing marked edge, and the cycle basis walk generally preserves the degree of the marked edges. A simple example where this symmetry is broken is given in \ref{sec:appendix}.

\section{Target Distribution}
\label{sec:target}

Recall that $\mathcal{X} = \{x = (T, M)\colon T \text{ is a spanning tree of } G, M \subseteq T\}$ is the lifted space. We define \begin{equation} 
P_d(G) = \{\xi\colon V \rightarrow \{1, 2, \ldots, d\} \mid \forall i, \xi_i \text{ is connected}\} ,
\end{equation} where $\xi_i$ represents the induced subgraph of $G$ on vertex set $\xi^{-1}(i)$. That is, $P_d(G)$ is the space of $d$-partitions of $G$ where each part is connected. Since MEW targets a distribution over $\mathcal{X}$, we choose a distribution $p(x)$ over $\mathcal{X}$ that induces a desired distribution $\pi(\xi)$ for $\xi \in P_d(G)$. 

\medskip

To achieve this, we choose a target distribution of the form \begin{equation}
    p(x) \propto \frac{\exp(J(\xi(x)))} {\tau(\xi(x))},
\end{equation} where $\xi(x)$ denotes the connected partition $\xi \in P_d(G)$ corresponding to the state $x \in \mathcal{X}$.  The degeneracy factor $\tau \colon P_d(G) \rightarrow \mathbb{Z}_+$ counts the number of states in $\mathcal{X}$ that map to the same partition in $P_d(G)$. The energy function $J \colon P_d(G) \rightarrow \mathbb{R}$ may be user-specified to incorporate various objective measures on $P_d(G)$. In this work, we include in every experiment a population constraint of 5\% that is wrapped into the energy function following \citet{autry2023metropolized}. 

\medskip

In this framework, multiple lifted states $x \in \mathcal{X}$ can correspond to the same partition $\xi \in P_d(G)$. The degeneracy factor \begin{equation}
    \tau(\xi) = \Pi_{i =1}^{d}t(\xi_i)\cdot t(Q),
\end{equation} accounts for this, where $t(\xi_i)$ counts the number of spanning trees in component $\xi_i$, and $Q = G \setminus \xi$ is the quotient graph. 

\medskip

It is important to note that while the degeneracy factor $\tau(\xi)$ corrects for the multiplicity of lifted states mapping to the same partition, it does not account for the distribution of $P_d(G)$ itself. Let $\pi(\xi) \propto f(E(\xi))$ be the target distribution, where $f$ is a desired function of an observable $E$. The induced marginal distribution over $E$ is $P(E) = f(E) \cdot \Omega(E)$, where $\Omega(E) = \sum_{\xi \in P_d(G)}\mathbbold{1}(E(\xi) = E)$ is the multiplicity of partitions with observable $E(\xi) = E$. Thus, we only recover $P(E) = f(E)$ when $\Omega(E)$ is (approximately) constant. \citet{DeFord2021Recombination} showed that there are far more partitions with higher cut edge counts than those with fewer cut edges, suggesting $\Omega(E)$ is not constant in this case. Therefore, when targeting a distribution based on cut edges as in Sections \ref{sec:compactnessI} and \ref{sec:Texas}, we may expect to see a bias towards less compact partitions. 

\medskip    

\begin{figure}[htbp]
    \centering

    \subfigure[]{\includegraphics[width=0.49\linewidth]{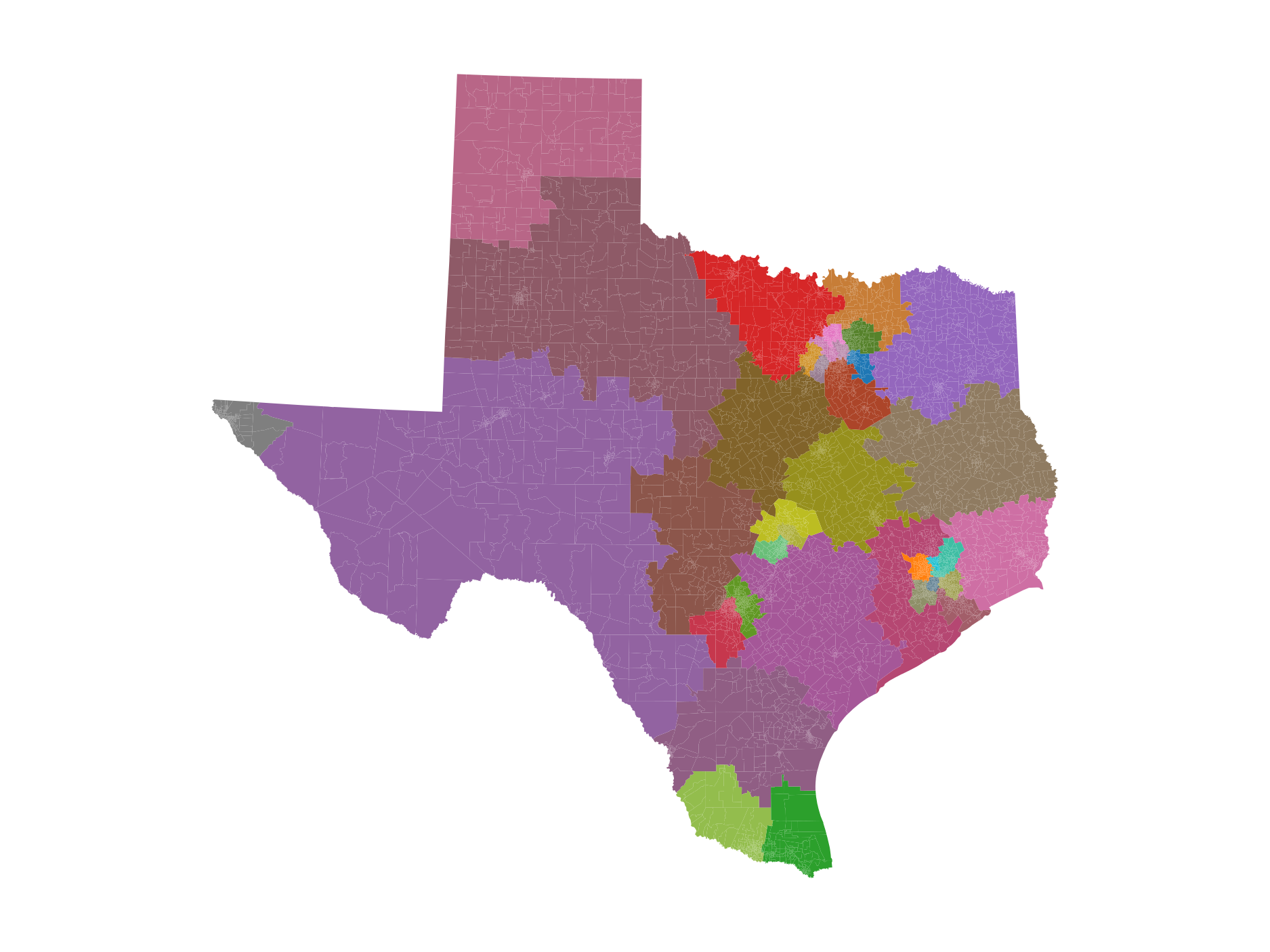}}
    \subfigure[]{\includegraphics[width=0.49\textwidth]{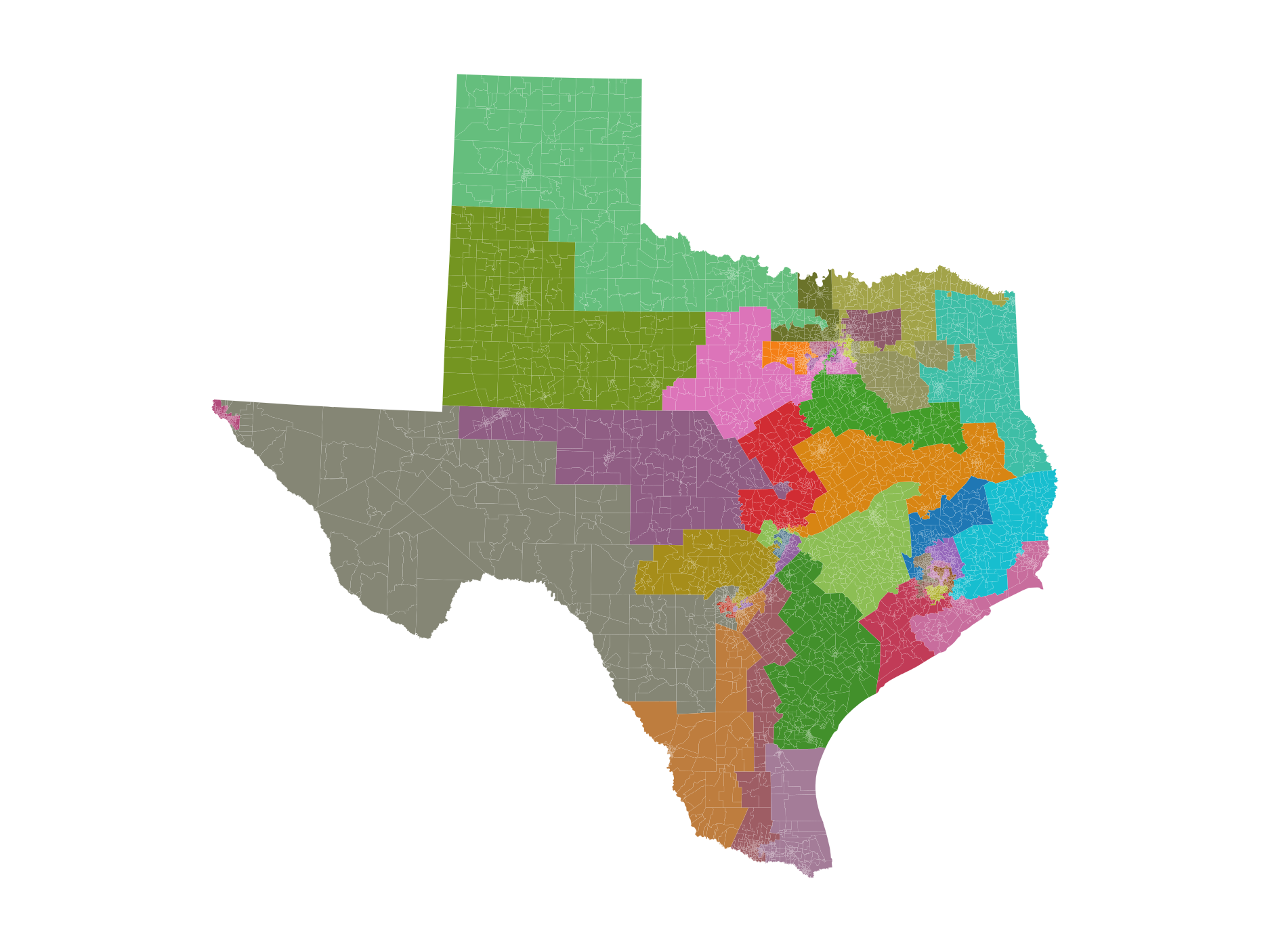}}
    \caption{Comparison of a sampled Texas redistricting plan (a) and the enacted 2020 plan (b). The spanning tree products differ by approximately $10^{466}$, demonstrating the extreme and impractical nature of spanning tree-based probability weighting in real-world applications.}
    \label{fig:tx_trees}

\end{figure}

This choice of target distribution allows us to sample in the lifted space $\mathcal{X}$ under a target distribution over $P_d(G)$. One reason this is so important is that, especially on large dual graphs, the spanning tree count of partitions is so astronomically large that it dwarfs other constraints that we might try to build into an energy function. For example, in Figure \ref{fig:tx_trees}, we show two possible redistricting plans of Texas's congressional districts. The map on the left is drawn from the uniform distribution on $\mathcal{X}$, corresponding to the spanning tree distribution, and the map on the right is the 2020 enacted plan. The sampled plan has a degeneracy factor of $\tau(\xi_{\text{sampled}}) \approx 10^{5160}$, and the enacted plan has a degeneracy factor of $\tau(\xi_{\text{enacted}}) \approx 10^{4694}$. Thus, when sampling from the spanning tree distribution, the sampled plan is about $10^{466}$ times more likely to be sampled than the enacted plan. To put the number $10^{466}$ into perspective: there are an estimated $10^{80}$ atoms in the observable universe. So, the sampled plan is more likely than the enacted plan by a factor that is astronomically larger than the number of atoms in the universe. This demonstrates that the spanning tree distribution gives enormous weight to highly compact plans, whereas enacted plans tend to have far more cut edges. This motivates the need for an algorithm that can shift the sampling distribution towards plans with a target number of cut edges, or more generally, encode policy-based criteria into the sampling process. 


\medskip

Previous studies discount the degeneracy factor \citep{autry2023metropolized, Autry_Multiscale}, taking a target of the form $p(x) \propto \exp(J(\xi(x))/\tau(\xi(x))^\gamma$, where $0 \leq \gamma < 1$. This only partially accounts for the spanning tree bias with the goal of achieving better mixing and convergence results. However, this partial correction still leaves the sampler heavily biased towards plans with large spanning tree counts. MEW's ability to maneuver locally in $\mathcal{X}$ allows us to use the full degeneracy factor ($\gamma = 1$), properly accounting for all configurations. This allows us to target a broader class of policy-based distributions over $P_d(G)$ without the overwhelming bias from spanning tree counts, allowing us to focus on other criteria such as compactness, competitiveness, or partisan symmetry measures, even when these criteria remain implicitly related to spanning trees.

\section{Results}
We evaluate the efficacy of MEW through numerical experiments on three real-world graphs with varying energy functions $J$. First, we target the uniform distribution over partitions of the dual graph of Cheshire County, an example that permits comparison to an enumerated baseline. Second, we target the spanning tree distribution on the dual graph of New Hampshire, and since ReCom draws independent samples from this distribution when $d = 2$, we can use it as a baseline. Third, we target competitiveness, compactness, and multivariate distributions over the dual graph of New Hampshire. Finally, we target a multivariate distribution on the dual graph of Texas's voting precincts to validate MEW's scalability to large dual graphs with many districts. Our results demonstrate strong evidence of mixing in all cases.

\subsection{Cheshire County}

\begin{figure}[htbp]
    \centering
    \includegraphics[width=0.7\linewidth]{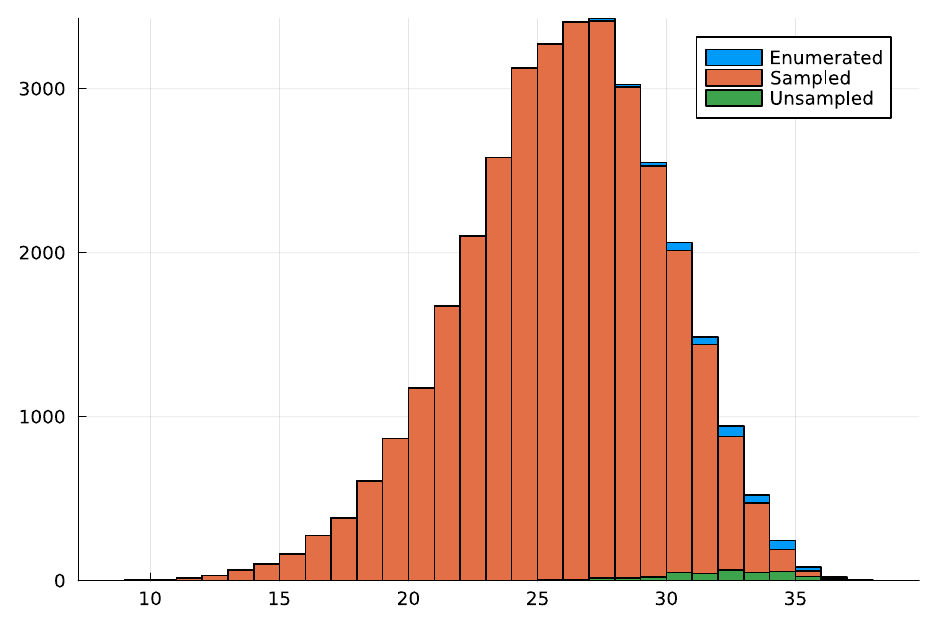}
    \caption{Comparison of cut edges for enumerated, sampled, and unsampled partitions in Cheshire County’s dual graph. After running MEW for 1 million steps, 99\% of the 34,225 balanced 2-partitions were sampled. Unsampled partitions (green) tend to have higher cut edges than sampled partitions (orange).}
    \label{fig:enumerated}
\end{figure}

The dual graph of Cheshire County, New Hampshire, has 27 vertices, 63 edges, and 34,225 balanced 2-partitions. By enumerating all the partitions in Cheshire County using the enumpart algorithm introduced in \citet{fifield2}, we can directly compare them with an ensemble generated by MEW. When we run MEW for 1 million steps targeting the uniform distribution on partitions ($J(\xi(x)) = \text{const.}$), we successfully sample ~99\% of the partitions. The cut edge counts of the sampled, unsampled, and enumerated partitions are shown in Figure \ref{fig:enumerated}. The unsampled partitions have more cut edges than the sampled plans. This is a similar result to the observed behavior of MFR \citep{autry2021metropolized}. After 10 million runs, MEW successfully samples all of the enumerated partitions, and the gap disappears. 

\medskip

What we observe here are the two opposing forces of MEW. One force is the force towards more compact plans. Since the cycle basis walk targets the uniform distribution on spanning trees, partitions that have higher spanning tree counts will be favored by the proposal step. Thus, we will systematically under-sample less compact plans. The opposite force pushes the walk towards less compact plans. The degeneracy factor $\tau$ in the target distribution encodes a higher acceptance probability for plans with lower spanning tree counts. When these two forces are in balance, the chain is successful. 

\subsection{The Spanning Tree Distribution}

We investigate the spanning tree distribution as a target distribution for two primary reasons. Since much attention has been devoted to its study, researchers may want to use MEW to target the distribution or related distributions. Secondly, since many algorithms are available to target the distribution, we can use it as another baseline to assess convergence. To this end, we employ ReCom, which, in the $d=2$ case, draws independent samples from the spanning tree distribution; each partition has probability proportional to the product of the spanning tree product of each part and the spanning tree product of the quotient graph. Notice that this is exactly the uniform distribution on $\mathcal{X}$, thus we set \begin{equation}J(\xi(x)) = \ln(\tau(\xi(x))).\end{equation} We note that this differs slightly from the definition of the spanning tree distribution given earlier, which omits the quotient graph term. However, this term is negligible in practice and does not materially affect the distribution.

\medskip

\begin{figure}[htbp]
    \centering
    \includegraphics[width=\linewidth]{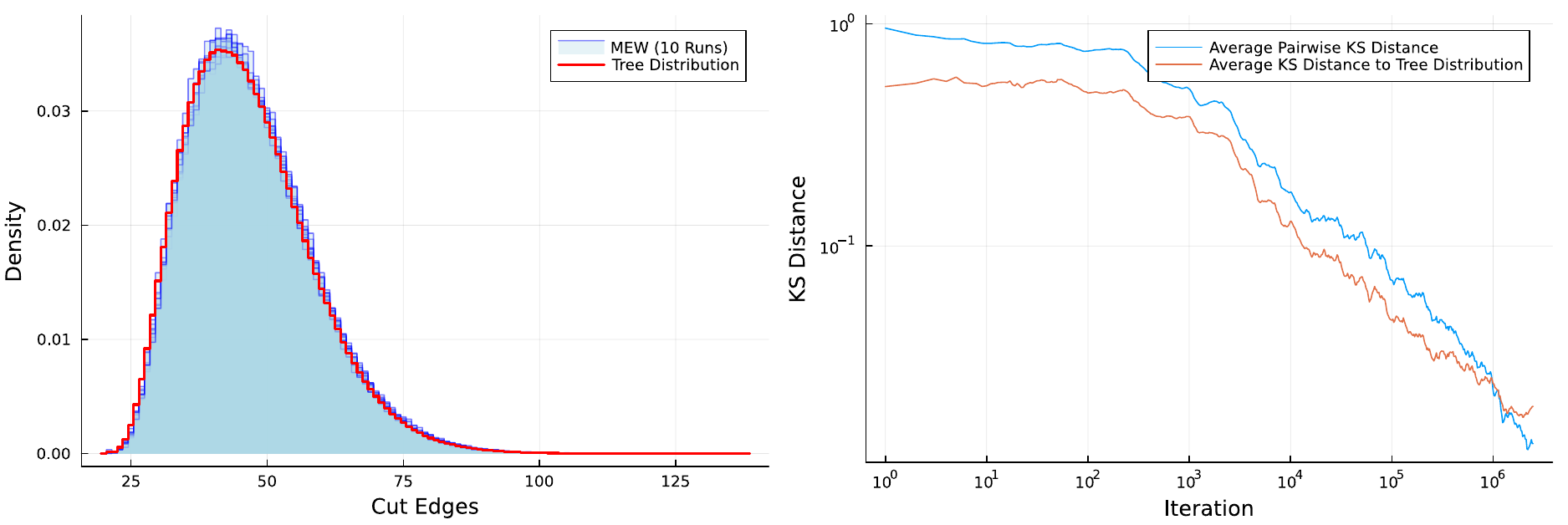}
    \caption{(a) Comparison of MEW (blue) to a Recombination baseline (red) when sampling with a target of the spanning tree distribution. (b) Decay of the Average Pairwise KS Distance (blue) and the Average Kolmogorov-Smirnov (KS) Distance to the Spanning Tree Distribution (red). After many iterations, MEW successfully samples from the spanning tree distribution, as evidenced by its convergence to the baseline.}
    \label{fig:spanning_tree}
\end{figure}

As we can see in Figure \ref{fig:spanning_tree}, MEW successfully samples from this distribution. The average pairwise Kolmogorov-Smirnov (KS) distance between the chains decays to nearly zero (0.0124 after 2.5 million steps), and the average KS distance from the chains to the baseline also decays to nearly zero (0.0188 after 2.5 million steps). The figure also reveals a limitation of sampling from the spanning tree distribution and related distributions. When sampling less compact partitions to target realistic plans (the enacted plan has 72 cut edges) or to investigate the effects of highly noncompact districts (targeting plans with 100 or more cut edges), we would see long mixing times and high rejection rates due to the distribution's naturally compact structure. 

\subsection{Competitiveness Distribution}
\label{sec:competitivenessI}
New Hampshire has two congressional districts, and previous investigations show that the seats are occupied by either two democrats or one democrat and one republican \citep{Palmer_Schneer_DeLuca_2024}. To ensure competitiveness in the second district, we construct an ensemble of redistricting plans with a tight race in the second district. We encode this in the energy function as \begin{equation}
    J(\xi(x)) = -10 \cdot (p - 0.5)^2,
\end{equation} where $p$ represents the share of democratic votes in the second district. Then the induced energy function on $P_2(G)$ is a normal distribution centered on 50\% with variance $1/20$. 

\begin{figure}[htbp]
    \centering
    \includegraphics[width=\linewidth]{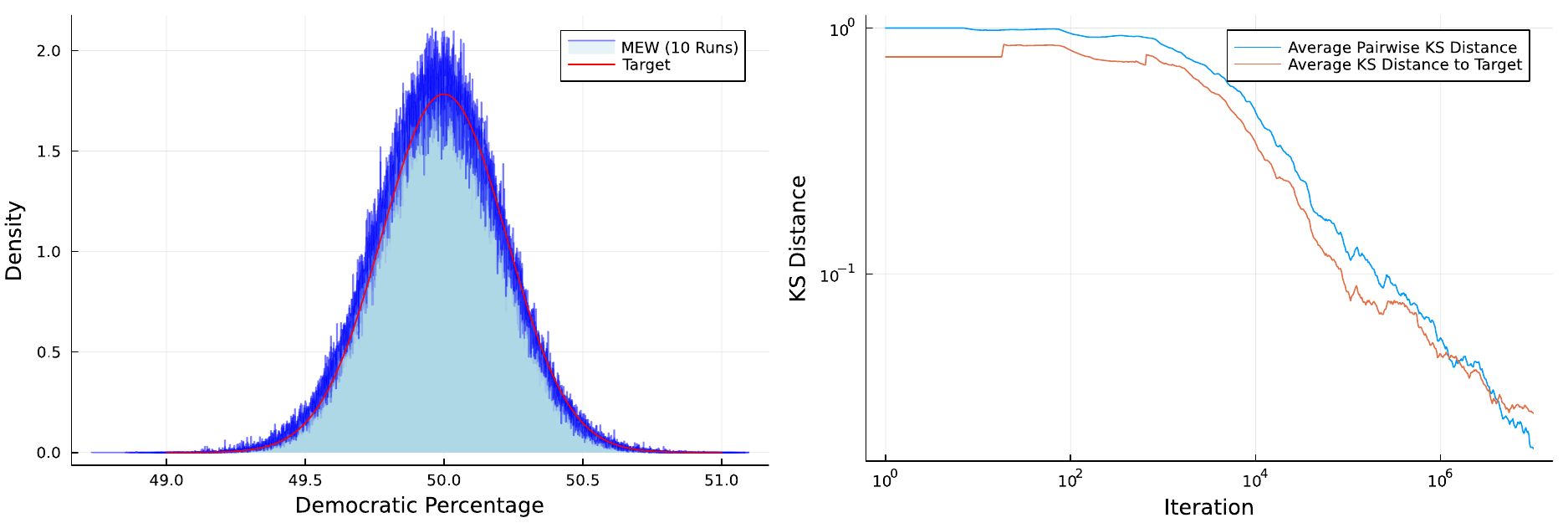}
    \caption{(a) Comparison of the sampled distribution (blue) with the target distribution (red), showing approximate agreement. (b) Average pairwise KS distance (blue) and average KS distance to the target distribution (red) vs iteration, showing decay to nearly zero after 10 million steps. These diagnostics suggest MEW is effectively sampling from the target distribution.}
    \label{fig:competitiveness_histogram}
\end{figure}

In Figure \ref{fig:competitiveness_histogram} (a), we plot the resultant distribution and can see that it roughly agrees with the target distribution. In Figure \ref{fig:competitiveness_histogram} (b), we plot the average pairwise KS distance and the average KS distance to the target distribution, showing that both decay to nearly zero after 10 million steps (0.019 and 0.027 respectively). Given these heuristics, we are confident that MEW is successfully sampling from this target distribution. 

\medskip

This is the first example in our experiments of targeting a Gaussian distribution. This is distinct from prior approaches, which typically use exponential energy functions that push the sampler towards more competitive or compact plans. By contrast, targeting a Gaussian centers the sampling window around a target value, such as 50\% Democratic vote share, and the resulting ensemble closely approximates the target when $\Omega(E)$ is approximately constant in the region of interest. This allows for a new ability to evaluate proposed legislative changes: by shifting the target distribution, one could assess the resulting partisan outcomes before any maps are drawn.

\subsection{Compactness Distribution}
\label{sec:compactnessI}

When sampling from the uniform distribution over graph partitions, \citet{najt2019complexity} showed that most plans in the space are extremely non-compact. To create a meaningful and realistic ensemble, we target compact plans.  The number of cut edges in a partition has become a popular metric for compactness in mathematical settings due to its low complexity and its relation to the border length of spanning forests \citep{clelland2021compactnessstatisticsspanningtree,Procaccia2021Compact}. To target realistic, compact plans, we assume an energy function \begin{equation}
    J(\xi(x)) = -0.1 \cdot (c - 72)^2,
\end{equation} where $c$ represents the number of cut edges. As before, this gives us a Gaussian centered on 72 with a variance of $5$. The value 72 corresponds to the number of cut edges in the enacted plan.

\begin{figure}[htbp]
    \centering
    \includegraphics[width=\linewidth]{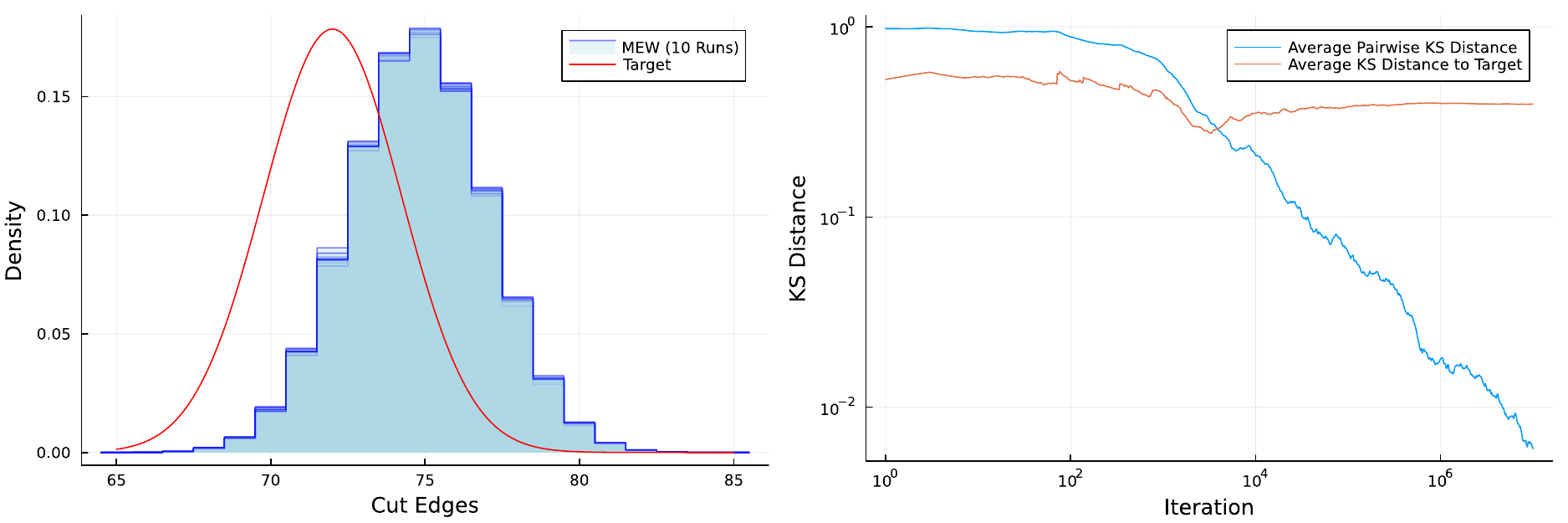}
    \caption{(a) Comparison of the sampled distribution (blue) against the target distribution (red). The empirical distribution is shifted from the target across all 10 chains. (b)  Average pairwise KS distance (blue) and average KS distance to the target distribution (red). The KS distance to the target plateaus whereas the pairwise KS distance decays to nearly zero, evidence of convergence to the shifted distribution.}
    \label{fig:compactness}
\end{figure}

\medskip

With this target distribution, we do not have the same tight agreement as before. In Figure \ref{fig:compactness} (a), we see that the empirical distribution is shifted relative to the target. However, this shifted distribution shows great evidence of convergence. In Figure \ref{fig:compactness} (b), we see that the pairwise KS distance decays to nearly zero. After 10 million steps, the average pairwise KS distance is 0.0061. 

\medskip

Following \citet{deford2019recombination}, we hypothesize that $P_2(G)$ is exponentially distributed with respect to cut edges in the region of interest for redistricting. This exponential distribution imposes a shift on the sampled distribution. In Appendix \ref{sec:appendix_experiment}, we conduct an experiment with a variable constructed specifically to have an exponential distribution and observe a similar shift behavior. 

\subsection{Multivariate Distribution}

In practical applications, researchers will need to encode various constraints into their energy functions. To investigate the behavior of the chain when sampling from a multivariate distribution, we may simply take a linear combination of the competitiveness and compactness energy functions. This yields a multivariate Gaussian with zero correlation. This energy function has four tunable parameters; the means control the center of the distribution, and the weights control the spread in each of the principal axes. A general energy function of this form is  \begin{equation}
    J(\xi(x)) = -\beta_1 (p - \mu_p)^2 - \beta_2 (c - \mu_c)^2,
    \label{eq: J}
\end{equation} where $\beta_1$ and $\beta_2$ are the weights on vote percentage and cut edges, respectively, $p$ represents the Democratic party's vote share in the second district, and $c$ is the number of cut edges. 

\medskip

\begin{figure}[htbp]
    \centering
    \includegraphics[width=\linewidth]{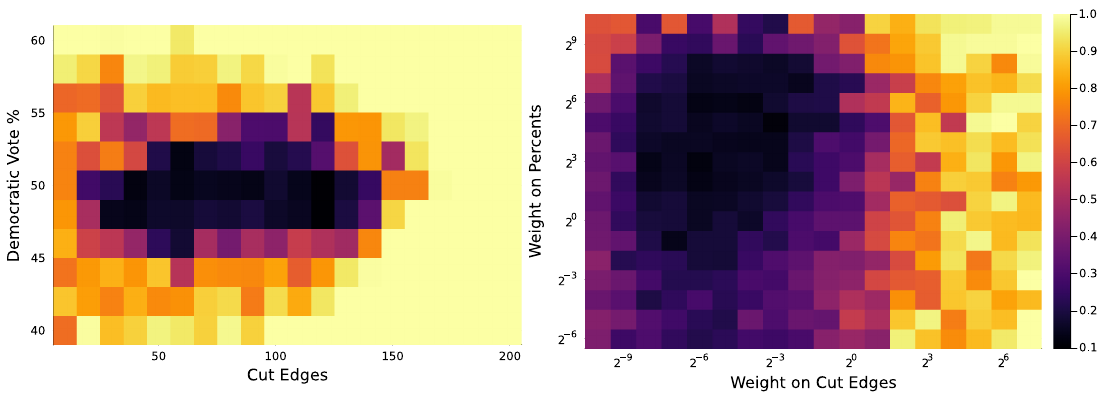}
    \caption{(a) Average pairwise KS distances as a function of target distribution means. The black region indicates the chain’s convergence zone after 100 thousand steps, centered on the practical region of interest (72 cut edges, 50\% democratic vote share). Poor convergence occurs outside this zone. (b) Average pairwise KS distances as a function of target weights, showing a broader convergence region. Convergence degrades sharply for cut edge weights $\beta_2>1$.}
    \label{fig:multi}
\end{figure}

In Figure \ref{fig:multi}, we plot the results of a convergence analysis designed to determine the parameter range within which MEW converges. 
In this analysis, 10 chains of 100 thousand steps are run at each parameter combination, and we plot the average pairwise (2-dimensional) KS distance in a heatmap. In Figure \ref{fig:multi} (a), we fix $\beta_1 = 10$ and $\beta_2 = 0.1$, and search a grid of target mean pairs. The black region represents a conservative estimate of the convergence region of the chain (since each chain only ran for 100 thousand steps). The convergence region covers the practical zone of interest, centered on 72 cut edges and 50\% democratic vote, and the worst convergence is observed well outside of this zone. 

\medskip

In Figure \ref{fig:multi} (b), we fix $\mu_p = 0.5$ and $\mu_c = 72$ and search a grid of target weight pairs. The convergence region is much larger than before, and we have a similar result to above; the convergence region contains the zone of practical interest. Rarely would researchers assign $\beta_2 > 1$, as this targets a very narrow distribution. Convergence falls off rapidly for cut edge weights larger than 1; any plan that differs from the target mean of cut edges will do so by several standard deviations. In the percentage axis, the convergence region is much wider.

\subsection{Texas}
\label{sec:Texas}

Many applications will be to states with much larger dual graphs and multiple districts. To test MEW's performance on such examples, we run the chain on the dual graph of Texas, which contains 8,933 vertices and 24,514 edges (compared to New Hampshire's 320 vertices and 854 edges). Again, we take a target distribution of a bivariate Gaussian on compactness and political metrics. 

\medskip

For compactness, we use the number of cut edges, and in the place of a competitiveness metric, we choose the Mean-Median score, a partisan symmetry metric. The score compares a party's average vote share to its median vote share across districts, with a positive score indicating an advantage for that party. While this metric has noted limitations \citep{DeFord2025Bounds,ratliff2025donttrustsinglegerrymandering}, as do all single-metric approaches, the ability to target explicit distributions on single metrics enables researchers to explore metric-specific implications in redistricting analysis. 

\medskip

With these two metrics, we target an energy function of the same form as Equation \ref{eq: J}, but now with $p$ representing the mean-median score, $\mu_p = 0$ (the mean-median score prescription for a perfectly `fair' plan), $\beta_1 = 100,000$, $\mu_c = 3,346$ (the enacted plan value), and $\beta_2 = 0.01$. 

\begin{figure}[htbp]
    \centering
    \includegraphics[width=\linewidth]{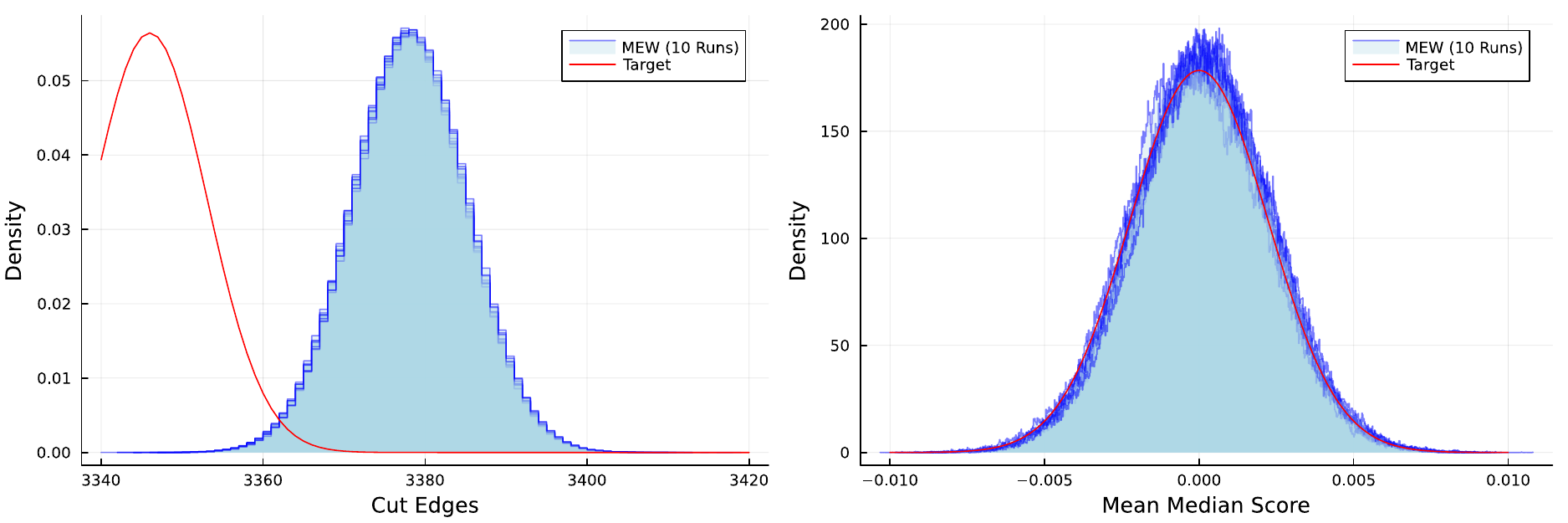}
    \caption{Marginal distributions for Texas. (a) Cut edges distribution displaying the characteristic shift observed in compactness metrics, centered at the enacted plan value of 3,346 cut edges. (b) Mean-Median score distribution showing convergence around the target value of 0.}
    \label{fig:TX}
\end{figure}

\medskip

Under this target distribution, we see good evidence of convergence. Over ten chains of 4 million steps each, the average pairwise (2-dimensional) KS distance between chains is 0.0285. In Figure \ref{fig:TX} we plot the marginal distributions. As in Section \ref{sec:compactnessI}, we observe a shift in the sampled marginal distribution of cut edges.

\section{Conclusion and Discussion}
\label{sec:discussion}

In this paper, we introduce the marked edge walk (MEW), a novel MCMC algorithm for sampling from the space of balanced graph partitions under a tuneable target distribution. The procedure combines a step of the cycle basis walk on spanning trees with a marked edge transition, and this balance of global and local steps allows for good exploration of the state space while maintaining easily calculable transition probabilities. Unlike previous approaches such as the Metropolized flip walk or Metropolized Forest ReCom (MFR), MEW mixes well under target distributions unrelated to the spanning tree distribution. 

\medskip

When targeting the uniform distribution on the dual graph of Cheshire County, the spanning tree distribution, and a competitiveness distribution on the dual graph of New Hampshire, the chain exhibits fast mixing evidenced by rapidly decaying pairwise KS distance and KS distance to the target distribution. 

\medskip

When targeting a compactness distribution based on cut edges, the empirical distribution is shifted from the target. However, the chains maintain strong convergence properties, with pairwise KS distances decaying to nearly zero. We hypothesize that this shift is due to the underlying exponential structure of $P_d(G)$. This hypothesis is supported by the observation of a similar shift observed on an observable with a known exponential distribution. MEW also converges well to a multivariate target distribution on the dual graph of New Hampshire and Texas, displaying a wide region of convergence and scalability to large graphs with many districts. 

\medskip

One limitation of MEW is computational efficiency. Each acceptance ratio includes calculating the spanning tree count of the districts, which due to Kirchoff's matrix-tree theorem is straightforward while computationally taxing. New Hampshire's dual graph contains only 320 vertices, while applications to other states (such as Texas with 8,933 vertices) see greatly increased run-times. The MFR algorithm includes memoization of this calculation to increase time efficiency. Additionally, the observed shift in the ensemble when targeting a compactness distribution suggests MEW favors plans that are slightly less compact than the target distribution. Future work on the algorithm could explore ways to correct for the bias, possibly by adjusting the target distribution or proposal step. Because we hypothesize that the shift is a result of the underlying structure of $P_d(G)$, research into quantifying this structure for different graph types would allow us to correct for the shift in the target distribution. 

\medskip

The ability to target a broader class of distributions opens possibilities for prospective redistricting analysis. The ensemble method has primarily been used to evaluate existing or proposed maps against a distribution of alternatives. However, MEW's capacity to sample from specified target distributions, particularly non-optimization distributions, allows for a new use case: evaluating the expected consequences of proposed redistricting criteria before maps are drawn. For example, policymakers considering the impact of population deviation strictness on partisan outcomes could target different population-based distributions and observe how the resulting ensembles differ in partisan outcomes. While such applications would require careful consideration of the biases we observe (such as the exponential tilt in Section \ref{sec:compactnessI}), they represent a potential shift from retrospective evaluation to prospective policy analysis. 

\medskip

And lastly, an important research direction is incorporating MEW in optimization schemes. Its calculable transition probabilities and favorable mixing make it a great candidate for tempering schemes like those used in \citet{Autry_Multiscale, chuang2024multiscaleparalleltemperingfast}. Because MEW can target a broader class of distributions, it could optimize for objectives that have been difficult to address with prior proposals. 



\section*{Supplementary Material}

The data and code needed to replicate these findings can be found at \url{https://github.com/amcwhorter/MEW/}. All computations were performed on doob, a 96-core 3.6GHz AMD Epyc system managed by the Dartmouth College Department of Mathematics.

\section*{Acknowledgments}
The authors thank the editor, associate editor, and the Journal of Data Science for considering our manuscript for publication. We also thank Kristopher Tapp and Greg Herschlag for helpful conversations.

\section*{Funding}
This work was funded by the Dartmouth Fellowship to Atticus McWhorter and by the Alfred P. Sloan Foundation (G-2021-16778) and the National Science Foundation (DMS-1928930) which supported Daryl DeFord during his visit to SLMath in 2023 where this collaboration began.

\bibliographystyle{elsarticle-harv}
\bibliography{Gerry_refs-3}

\begin{appendix}

\setcounter{figure}{0}
\section{Asymmetric Transition Probability}
\label{sec:appendix}

\begin{figure}[htbp]
    \centering
    \includegraphics[width=0.5\linewidth]{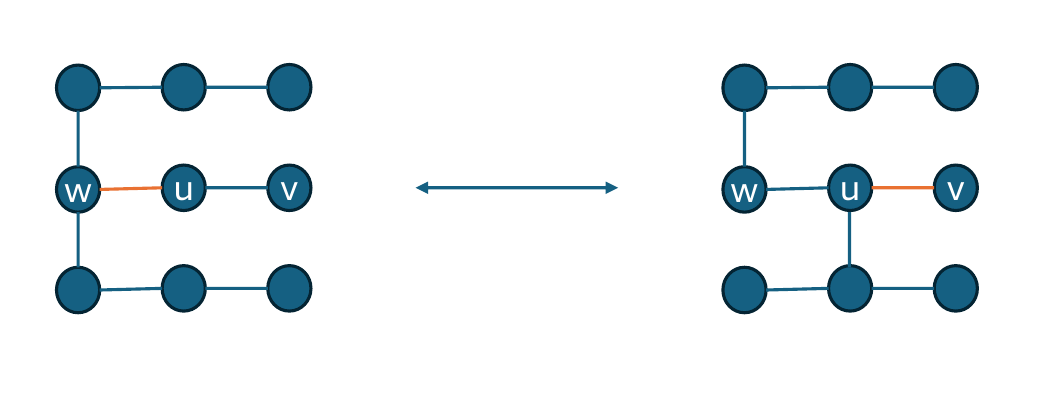}
    \caption{Example states $x = (T, M), x' = (T', M')$ that breaks symmetry}
    \label{fig:A1}
\end{figure}

Here, we include an example transition that breaks symmetry from a state $x = (T, M)$ (Figure \ref{fig:A1}, left) to $x' = (T', M')$ (Figure \ref{fig:A1}, right). In this case, the forward transition probability, $P(M'|M, T') \propto 1/\deg_{T'}(u) = 1/3$. In the reverse transition, $P(M|M', T) \propto 1/\deg_T(u) = 1/2$. Since the neighborhood of $u$ changes between $T$ and $T'$, the forward and reverse transition probabilities are not the same.

\section{Exponential Shift Experiment}
\label{sec:appendix_experiment}

Because of the unknown structure of $P_d(G)$, there is no guarantee that the sampled distribution of an observable will show perfect agreement with the target distribution. In this example, we hypothesize that the underlying distribution of $P_2(G)$ is exponentially distributed with respect to cut edges, as discussed in \citet{DeFord2021Recombination}. 

\medskip

To investigate this hypothesis, we first consider a toy model. Consider a Markov chain that draws independent samples from an exponential distribution $g(x) = \lambda \cdot\text{exp}(-\lambda x)$ and uses Metropolis-Hastings rejection to target a Gaussian distribution of the form $p(x) \propto \text{exp}(-\beta(x - \mu)^2)$. The acceptance ratio would include the ratio of the two distributions, which simplifies to a shifted Gaussian with mean $\hat\mu = \mu + \lambda/2\beta$ and variance $\hat\sigma^2 =1/2\beta$. If we do not take into account the effect of the underlying exponential distribution, we are effectively targeting an exponentially tilted distribution with mean $\hat\mu = \mu - \lambda/2 \beta$ and variance $\hat\sigma^2 = 1/2\beta$. If our hypothesis were true, when sampling from an underlying exponential distribution with parameter $\lambda$, we expect a shift $\partial\mu = \hat\mu - \mu = \lambda/2 \beta$.

\medskip

We then construct an observable with a known exponential distribution. We assign each node $n \in V$ a weight $w_n$ drawn independently and uniformly from $[0, 1]$. Let $\mathbf{s}$ denote the vector of total weights summed across parts. For sufficiently large parts, the central limit theorem ensures each component $\mathbf{s}_i \sim \text{N}(|\xi_i|/2, |\xi_i|/12)$, where $|\xi_i|$ denotes the number of nodes in $\xi_i$. 

\medskip

Without loss of generality, we focus on $\mathbf{s}_1$. To construct an exponentially distributed variable $Z \sim \text{Exp}(\lambda)$, we apply a probability integral transform. First, we standardize $\mathbf{s}_1$ and apply its CDF. \begin{equation}
    U = \Phi\left( \frac{\mathbf{s}_1 - |\xi_1|/2}{\sqrt{|\xi_1|/12}}\right),
\end{equation} where $\Phi$ is the standard normal CDF. Since $U \sim \text{Uniform}(0, 1)$, we then apply the inverse transform \begin{equation}
    Z = -\frac{1}{\lambda}\ln(1 - U)\sim \text{Exp}(\lambda). 
\end{equation}

\medskip

With the exponentially distributed $Z$ as our variable of interest, we run 10 chains of 1 million steps each, targeting a normal distribution with mean $\mu = 2$ and variance $\sigma^2 = 1/2\beta$. Recall that our toy model predicts that we will converge to a Gaussian with mean $\hat\mu = \mu - \lambda/2\beta$ and variance $\hat\sigma^2 = 1/2\beta$. First, fixing $\lambda = 1$ and varying $\beta$, we observe a shift $\partial\mu \approx -0.57 (\beta^{-1}) + 0.01$ with correlation coefficient $r = - 0.97$. Additionally, the sampled variance satisfies $\hat\sigma^2 \approx 0.49 \beta^{-1}$ with correlation coefficient $r = 0.999$. When varying the exponential parameter (with fixed $\beta = 32$), we find $\partial\mu \approx-0.87 (\lambda/\beta) + 0.01$ with correlation coefficient $r = -0.97$ while $\lambda$ is uncorrelated with the sampled variance. 

\medskip

Overall, this experiment demonstrates that MEW exhibits systematic shifts when sampling over an exponentially distributed observable, and the behavior closely resembles that predicted by rejection sampling of independent draws from an exponential distribution. While the empirical coefficients deviate slightly from theoretical predictions, this supports the hypothesis that the underlying distribution of $P_d(G)$ with respect to cut edges is approximately exponential.

\end{appendix}

\end{document}